\documentstyle[twoside,fleqn,espcrc2x,epsf]{article}

\newcommand{\AmS}{{\protect\the\textfont2
  A\kern-.1667em\lower.5ex\hbox{M}\kern-.125emS}}
\def\makeadmark#1{\hbox{$^{\rm #1}$}}
\def\simlt{\hbox{ \rlap{\raise 0.425ex\hbox{$<$}}\lower 0.65ex\hbox{$\sim$} }}
\def\simgt{\hbox{ \rlap{\raise 0.425ex\hbox{$>$}}\lower 0.65ex\hbox{$\sim$} }}

\def\apj{{\rm ApJ}}

\def\etal{{\it et al.}}

\title{ Search for Earth Mass Planets and Dark Matter Too}

\author{S.H.Rhie\rlap,\makeadmark{a}  and
D.P.Bennett\rlap,\makeadmark{a,b,c,d}
\address{Lawrence Livermore National Laboratory, Livermore, CA 94550}
\address{Center for Particle Astrophysics, University of California,
    Berkeley, CA 94720}
\address{Department of Physics, University of California, Davis, CA 95616}
\address{Department of Physics, University of Notre Dame, Notre Dame, IN 46556}
\thanks{Work performed at LLNL is
supported by the DOE under contract W-7405-ENG.  Work performed by the
CfPA personnel is supported by the NSF through AST 9120005.} 
}

\begin{document}

\begin{abstract}
Gravitational microlensing is known for baryonic dark matter searches.  
Here we show that microlensing also provides  a unique tool for the 
detection of low mass planets (such as earths and neptunes) from the ground.  
A planetary system forms a binary lens (or, a multi-point lens),
and  we can determine the mass ratio of the planet with respect to
the star  and relative distance ($=$ separation/Einstein ring radius)
between the star and planet.   Such a microlenisng planet search project 
requires a $\approx 2$ m survey telescope, and a network of $1.5-2$ m follow-up 
telescopes capable of monitoring stars in the Bulge on a 24-hour basis.   
During the off-season of the Galactic bulge, this network can be used 
for dark matter search by monitoring the stars in the LMC and SMC.

\end{abstract}

\maketitle

\section{INTRODUCTION}

Dark matter searches are hybrids of particle physics and astrophysics
in many aspects,  and naturally,  one can find  the  infusion of 
techniques and research paradigms from  one field  into the other.   
One very important fallout of this hybridization is that the importance  
of single purpose dedicated telescopes is being widely recognized.  
Baryonic dark matter searches utilize  gravitational microlensing
phenomenon  that  occurs when a foreground dark object traverses
the line of sight of a target star (in the LMC, SMC and M31).  
What is measured in the microlensing experiments  is the {\it time
variation}  of the brightness of the target stars.   Therefore the stars
have to be monitored constantly and that is not unlike tending an
accelerator beam line day and night.  (A subtle difference must be that
an observer does not have to stay up during the day.)
This ``new mode" of telescope usage is orthogonal (and complementary) to the
conventional ``time-sharing mode"  where ``photon-starved" astronomers build
the largest possible telescopes the funding allows and share them 
by allocating each observer a few nights at  a time so that many different
astrophysical phenomena are pursued independently and thus incoherently.

This ``new telescope mode"   is     
expected to become a new tradition in astronomy  due to the enormous 
success of the current microlensing experiments (See Pratt {\it et al},  
Bennett {\it et al}  and Lehner {\it et al} in this volume).    
It is not hard  to imagine the importance of single 
purpose dedicated telescopes  if one is interested in learning the 
{\it dynamics} of the celestial bodies {\it directly} by monitoring their 
{\it changes in time}.       
The impressive catalog of $\sim 40,000$ variable stars collected by the
MACHO experiment   should be only the beginning  
of what are to come in the near future.   
The bread-and-butter astrodynamics
single purpose telescopes can bring is boundless.   
Here we  advocate that one of the most immediate
beneficiaries of the ``new tradition" of dedicated telescopes
in optical and infrared bands  
can be the search for low mass planets through microlensing.

The trademark of microlensing signature of baryonic dark matter has been
advocated to be the light curves that are achromatic, symmetric and
non-repetitive.  The symmetric light
curve is due  to the spherically symmetric gravitational potential of
a point mass.    When the mass has a small companion, the distribution
of the gravitational potential changes ever slightly, but the light curve
can have a substantial deviation from the symmetric shape, even though
briefly in time.    That is because lensing is a catastrophic phenomenon.
We can capitalize on this catastrophic behavior of
lenisng to detect planets as small as the earth.
Microlensing is the only ground-based method capable of detecting earth-mass
planets and providing planetary statistics much needed for the future
space-based planet search programs envisioned by the ExNPS panel.

\begin{figure}[thbp]
\begin{center}
\leavevmode
\hbox{%
\epsfxsize=7.9cm
\epsffile{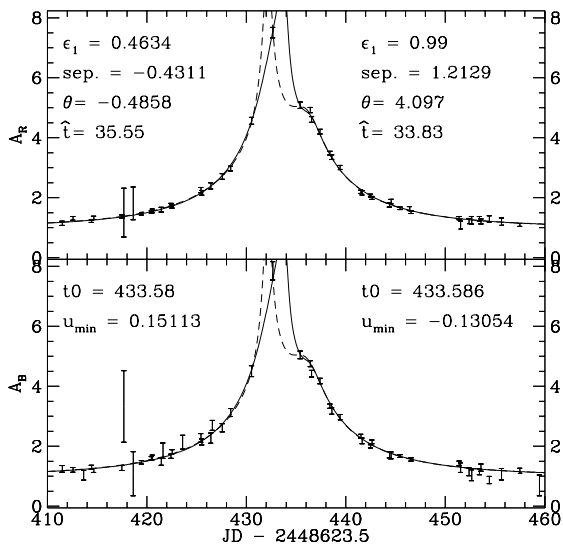}}
\vspace{-0.65cm}
\end{center}
\vspace{-1.2cm}
\caption {The binary lens fits to the GPE:  The solid curve is the small
mass fit for $\epsilon = 1 - \epsilon_1 = 0.01$, and the lens parameters
are shown on the right.   The dashed curve is the small
separation fit for $\epsilon = \epsilon_1$, whose lens parameters are shown
on the left.  $\hat t$ is the transit time
of the Einstein ring radius of the total mass, $|sep|$ is the separation,
$\theta$ is the angle of the source trajectory with respect to the lens
axis where $\epsilon_1$ is the fractional mass to the positive direction
in our convention, and $|u_{\rm min}|$ is the distance of the source
trajectory to the lens axis.}
\label{dm96_fig1}
\end{figure}

\section{The First Planet found through Microlensing?}

The first candicate Jupiter mass planet was ``found" in the first 
microlensing candidate event (GPE: Gold Plated Event) of the MACHO experiment.
The microlensing event toward the LMC is achromatic and fits
the symmetric single lens light curve with peak amplification $7.2$
fairly well and there is no doubt that that is a microlensing event  
(Alcock \etal, 1993). However,  there are a couple of ``anomalous data 
points" near the peak that are systematically shifted to the left
from the single lens fit curve.  One of us found that the `misfit'
can be explained  if the lensing object has a companion of the fractional 
mass $\epsilon=0.01$ (Rhie, 1994).   On the other hand,  the assumption of 
a binary lens introduces three more fit parameters and that leaves room 
for other fit parameter values.  
Dominick and Hirshfeld (1994)
found that the data can  be fit with $\epsilon = 0.414$.   
A binary lens system is characterized by two parameters, namely, the 
mass ratio  and the separation (tranverse distance on the lens plane)
between the two masses, and 
a binary lens behaves  very closely to a single lens (of 
the total mass) when the mass ratio or the separation is small.   
In other words,  when the GPE is considered as a binary
lensing event,  the parameter space is confined to two small regions of 
small mass ratio  or small separation because of the close proximity to the
single lens behavior.  (The binary lenses in the parameter space 
 of small mass {\it and} small 
separation  are practically  a single  lens  and  can not explain 
the ``deviation points".)    Our best fit values are $\epsilon=0.01$
and $\epsilon=0.463$ and the corresponding fit curves  are shown 
in figure~\ref{dm96_fig1}.   
One should note that the two fit curves are quite
distinct: The highest amplification point lies  on the rising side 
of the curve with $\epsilon =0.01$  and on the falling side of the curve with 
$\epsilon = 0.463$.   In retrospect,  one more data point near the peak 
might have resolved the dilemma of whether the lensing star has a Jupiter
mass planet or  is a dwarf binary star.   For this particular event, 
we wouldn't ever know because the probability for the lensing object to 
lens another star is $\sim R_{E}^2 n \sim 10^{-6}$, where $R_{E}$ is the
Einstein ring radius and $n$ is the surface number density of the LMC.

\section {Microlensing Signature of Earth Mass Planets}

If we summarize the lesson from the GPE event as a binary lens,
 $(1)$ \ When the mass of the companion is small ($\epsilon << 1$), 
the microlensing light curve is largely that of a single lens.
\ $(2)$ \  However, the small mass companion can produce an unmistakable  
signal by modulating the single lens curve substantially. 
\ $(3)$ \  The modulation signal lasts only briefly and  the unambiguous
detection can be made only through dense sampling of the light curve.  
\ In addition, \ $(4)$ \  The separation (transverse distance between
the star and planet on the lens plane) can not be too big because the
planetary signal will be dissociated from the stellar signal.
Therefore,  the separation has to be within a certain interval,  
and the interval is called the `lensing zone'.    Of course, the 
``lensing zone" depends on the sensitivity of the detectors,  and it
will turn out to be $\approx 0.6 R_{\rm E} - 1.6 R_{\rm E}$ for low
mass planets, where $R_{\rm E}$ is the Einstein ring radius of the total 
mass ($\approx$ stellar mass).   What should be noted here is that the 
``lensing zone" scales with the Einstein ring radius $\propto \sqrt{
\rm stellar\ mass}$.  It is a practical ``rule of thumb" that
the ``lensing zone"  is given by $\approx a^{-1}R_{\rm E} - aR_{\rm E}$, 
where $a (> 1)$ is the fudge factor depending on the mass ratio of the 
planet, detection strategy, and etc.

The duration of a microlensing event depends on
many parameters such as the mass, transverse velocity and reduced distance
of the lens, and also the size of the source star.  
Howevever,  for the current microlensing experiments toward  
the Galactic Bulge and the Large Magellanic Cloud,  one can estimate  
the duration as a function of the mass of the lens by considering 
the typical transverse velocity and reduced distance.
The mass dependence goes as $\propto \sqrt{{\rm mass}}$,  and 
the duration for a solar mass object is typically a couple of months
and a few days for a Jupiter mass brown dwarf, etc.   
When the Jupiter mass object is a planet around a star, we can 
estimate that the modulation duration is typically  a few days.    
If we consider the  exposure time of a few minutes as is the case in 
the current microlensing experiments toward the
Bulge,   one can sample a given
modulation due to a Jupiter mass planet  about 1400 times in principle.
Of course,  the currently active telescopes for microlensing experiments 
are survey telescopes and can not afford to follow one event with such a 
scrutiny, but it demonstrates that planet search via microlensing is not 
an idle idea at all.       

What one immediately realizes is that earth mass planet search via 
microlensing is also a sure possibility.  The modulation duration due to 
an earth mass planet is a few hours and hence the signal can be sampled 
as many as 45 times in principle.   The modulation signal typically has
the shape of a {\it wavelet} and  one can resolve the peaks and 
troughs  of the ``modulation wavelet" without ambiguity.    Actually, 
detecting Mars's ($0.1 m_{\oplus}$) dosn't seem to be an impossibility 
when estimated  along the same line.    However, the size of the source
stars begins to affect the signal seriously  when the mass of the planet
falls below 10 $m_{\oplus}$ or so.    In other words, the size of the 
source star (precisely speaking, the size of the star projected onto the 
lens plane with the observer at the focus of the projection)  becomes
comparable to the variation in the modulation wavelet (of a point source),
and the planetary signal gets smoothed over due to the integration effect.  
If the modulation wavelet has comparable troughs and peaks,  the planetary
signal can be completely wahsed out,   whose case we would like to term 
``interference effect".   If the peak is dominant over the troughs -- or,
the trough is dominant over the peaks,   the 
signal will not be averaged to zero, but it gets 
broadened and eventually  buried  below the measurement error, which 
we may term ``broadening effect" (meaning smoothing without destructive
interference).   

Therefore, it is important to carry out realistic calculations to
decide upon the feasibility of detecting $1 - 10 m_{\oplus}$ planets.
According to our calculations (Bennett and Rhie, 1996),
the probability to detect the planetary signal  of
an $10 m_{\oplus}$ planet in the lensing zone is $\sim 15\%$,
and for an earth mass planet, the probability drops to
$\sim 2\%$.   We expect that the probability will decrease even more
drastically for the mars mass planets, and a work is in progress for
the sake of confirmation.

\begin{figure}[thbp]
\begin{center}
\leavevmode
\hbox{%
\epsfxsize=7.9cm
\epsffile{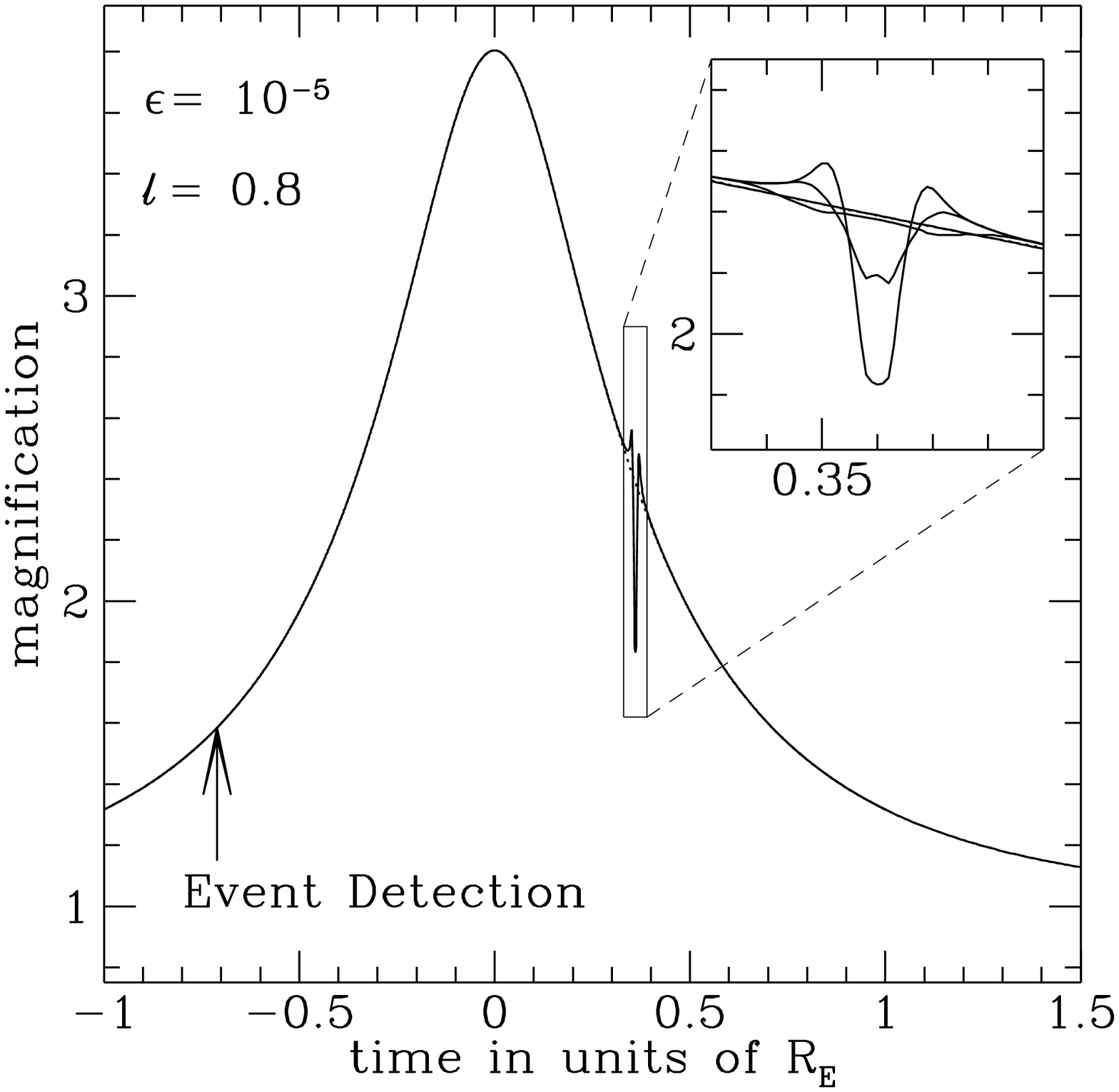}}
\vspace{-0.6cm}
\hbox{%
\epsfxsize=7.9cm
\epsffile{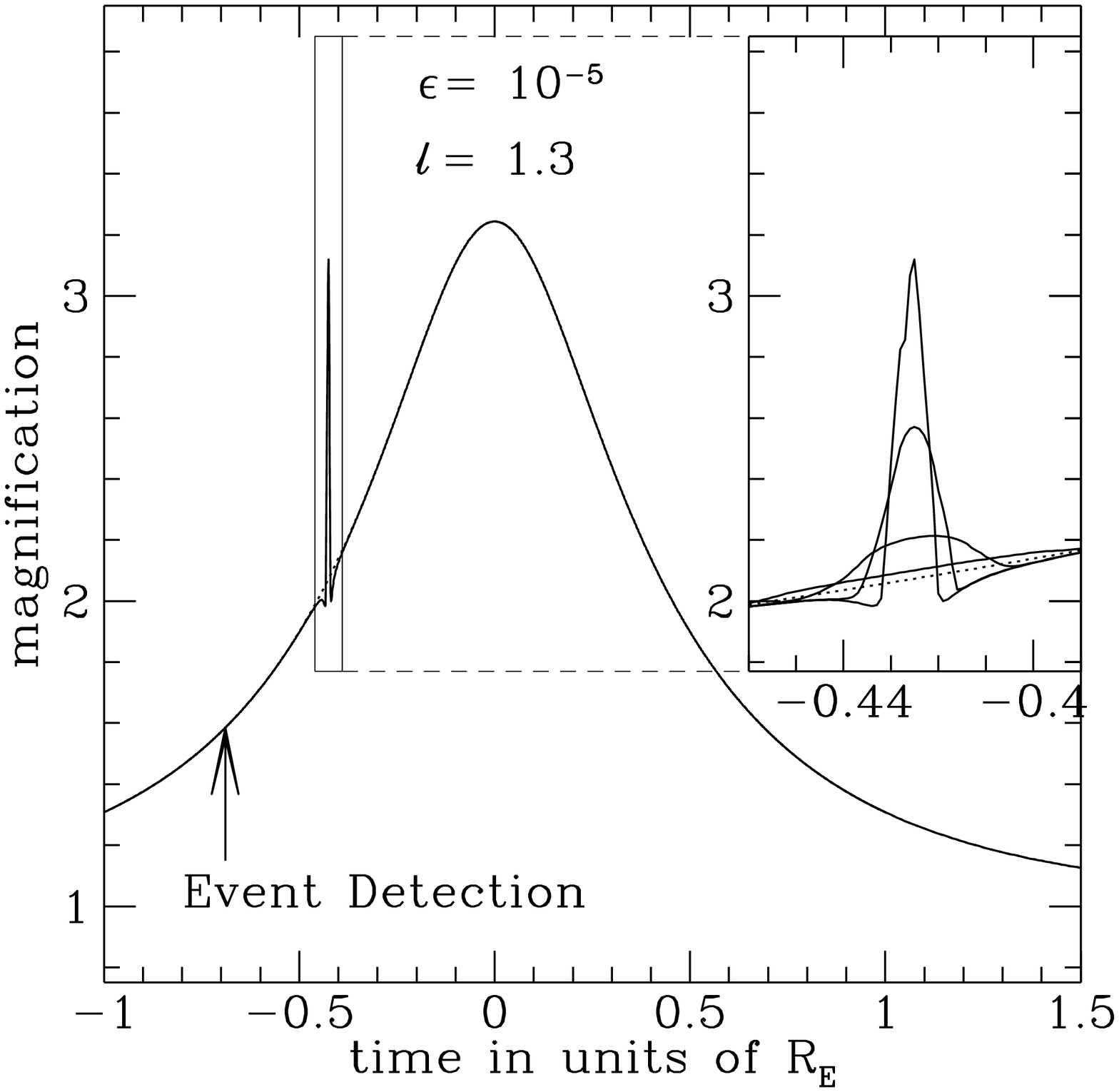}}

\end{center}

\caption {The microlensing signature of earth mass planets orbiting  
stars of mass $0.3 M_{\odot}$ in the Galactic disk toward the Bulge
with separations $\ell = 0.8$ (upper panel) and $\ell = 1.3$ (lower panel).  
The main plots are for a stellar radius of $r_s = 0.003$ while the insets
show the effect of increasing stellar size 
($r_s =$ 0.003, 0.006, 0.013 and 0.03). }
\label{dm96_fig2}
\end{figure}

\section {Planetary Binary Lenses}

In order to discuss what we can learn from a given planetary binary
lensing event,  it is necessary to know about binary lenses.  
When a planetary system falls in the line of sight of a background star,
the planetary system can be considered primarily as a binary lens 
because what matters is the planet falling in the lensing zone.
(Of course, more than one planets can fall into the ``lensing zone", 
and the signature will be multiple ``modulation wavelets".)
A binary lens simply means that the lens system consists of two bodies
jointly governing the gravitational potential that determines the optical
paths,  and the resulting configurational behavior of the images and
their sources are described by the binary lens equation.   
If $\omega$ and $z$ denote the source and image
positions in the lens plane as a complex plane, the binary lens equation is
\begin{equation}
\label{eq-bilens}
    \omega  = z - {1-\epsilon\over \bar z - \bar x_s}
                - {\epsilon\over \bar z - \bar x_p} \ ,
\end{equation}
where $\epsilon$ is the fractional mass of the planet, and $x_s$ and $x_p$
are the positions of the star and planet respectively. We work in units
of the Einstein radius, $R_E$, of the total mass $M$. Eq.~(\ref{eq-bilens})
has 3 or 5 solutions ($z$) for a given source location, $\omega$.

If $J_i$ is the Jacobian determinant of (the transformation given by) 
the lens equation at the position of the $i$-th image, 
the amplification of the image is 
given by the size of the image with respect to the size of the source. 
Therefore, the microlensing amplification (or total amplification) is given 
by 
\begin{equation}
\label{eq-Asum}
  A = \sum_i |J_i|^{-1}   \ . 
\end{equation}
The sign of $J$ describes the relative orientation of the area elements, 
and hence an image with $J<0$ is a flipped image and an image with 
$J>0$ is an unflipped image.  $J=0$  not only defines the boundary 
between flipped and unflipped images but also where the images enormously
brightens because of the inverse relation with the amplification $A$.
This singular (or catastrophic) behavior is at the heart of the 
microlensing as one of the most powerful tools in planet search.
The source positions that produce images falling on the critical curve
($J=0$) are called  {\it caustic curve},  and the caustic curve of the 
binary lenses changes its shape, size, position, and topology by joining
and splitting as the lens parameter 
($\epsilon$ and $l \equiv |x_s - x_p|$) changes.  
The caustic of a binary lens consists of one, two, or three 
{\it closed cuspy loops}, and this geometric diversity needs to be
understood somewhat rigorously because that is where the planetary
signature lies.  

For a planetary binary lens ($\epsilon << 1$), one caustic (almost a point) 
is always near the star (just as in a single lens -- so, we might as well
call ``stellar caustic"),  and hence the planetary signature ``modulation
wavelets" are determined by the distribution of the ``planetary caustics".
When the separation $l >  1$,  there is one  ``planetary 
caustic"  (of  diamond shape)  near the planet, and the 
``modulation wavelets"  are ``peak-dominated".  When the separation  
is $l > 1$,  there are two triangular shape  caustics (reflection symmetric), 
and the ``modulation wavelets"  tend to be ``trough-dominated". 
(See the plates in Bennett and Rhie, 1996.) \

\section {Conclusion}

So,  what can we learn about the planet once a planetary microlensing
event is detected?   As we have discussed before, the mass ratio of the 
planet with respect to the stellar mass can be determined without ambiguity.
The projected distance between the star and the planet can be determined
(in units of the Einstein ring radius) from the time difference between
the peak of the stellar light curve and the appearance of the planetary
signal.   The possible two-fold ambituity -- the separation $l$ or
$l^{-1}$ ($l > 1$) -- can be resolved because of the distinctive nature
of the modulation wavelets for $l > 1$  and $l < 1$.  

More than one planet can fall into the ``lensing zone",  and the lens 
system  may have to be considered as an $n$-point lens system where
$n > 2$.  However,  the  gravitational interference between the planets
can be ignored  most of the time and  hence the signature of two planets,
for example, in the ``lensing zone" will be simply two modulation wavelets 
on the symmetric stellar light curve.   (The interference  becomes 
singnificant  when the separation of the planets is order of the Einstein
ring radius of the total mass of the planets.)    

The only feasible target site for microlensing planet search is the Galactic
Bulge not only because the other galaxies have low microlensing probabilities
(the detection rate by the MACHO experiment toward
the LMC is $3-4$ per year) but  because  one is looking for lensing
evnets by normal stars (main sequence stars) that may host planets. 
 With a $2$m survey telescope, one can detect  
$\approx 250$ microlenisng events,  and a network of $1.5 -2$ m follow-up 
telescopes in Australia, Chile and the South Africa  (and also the 
South Pole) will be able to monitor the events with sufficient precision. 
With photometric precision $1\%$, the detection probability of an earth
mass planet in the lensing zone is $\approx 2\%$ and hence one may be
able to detect a couple of earth mass events per year.   The  frequency
of planets and especially that of earth mass planets constitutes a totally
unknown territory.    It is an exciting possibility that one can detect
planets {\it unambiguously} through microlensing from earth mass to 
Jupiter mass.    

   The Galactic Bulge is not visible during the austral summer (Nov., Dec.,
Jan.), and the telescope network can be pointed toward the satellite
galaxies for dark matter search or  possible signatures of planets 
orbiting the stars in the satellite galaxies.

\end{document}